\documentclass[aps,pre,reprint, amsmath, amssymb,superscriptaddress]{revtex4-1}

\usepackage{morefloats}
\usepackage{bm}
\newcommand{\beq}{\begin{equation}}
\newcommand{\eeq}{\end{equation}}

\usepackage[retainorgcmds]{IEEEtrantools}
\usepackage{graphicx,tikz,placeins}
\usepackage{mathrsfs}
\usepackage{amsmath,amssymb,amsfonts,physics}
\usepackage{color}
\usepackage{float}
\usepackage{times,txfonts}
\usepackage{nicefrac}
\usepackage{verbatim}

\usepackage[colorlinks=true,linkcolor=blue,urlcolor=blue,citecolor=blue,pdfusetitle]{hyperref}
\usepackage{physics}
\tikzset{d/.style={minimum width=7pt,inner sep=0pt,circle,fill=black}}
\usepackage{soul}
\newcommand{\mom}[1]{\left\langle#1\right\rangle}

\begin{document}
\title{Thermodynamics of interacting systems: the role  of the topology and collective effects}

\author{Iago N. Mamede}
\affiliation{Universidade de São Paulo,
Instituto de Física,
Rua do Matão, 1371, 05508-090
São Paulo, SP, Brazil}
\affiliation{Niels Bohr Institute, University of Copenhagen, Blegdamsvej 17, Copenhagen, Denmark}
\author{Karel Proesmans}
\affiliation{Niels Bohr Institute, University of Copenhagen, Blegdamsvej 17, Copenhagen, Denmark}
\author{Carlos E. Fiore}
\affiliation{Universidade de São Paulo,
Instituto de Física,
Rua do Matão, 1371, 05508-090
São Paulo, SP, Brazil}
\date{\today}

\begin{abstract} 
%Collective effects emerging from interacting units have been claimed as  powerful candidates for enhancing the power and efficiency of heat engines. 
%However, very little is known about the topology (network) of interaction interplays and their influences over the system performance. 
We will study a class of system composed of interacting
{unicyclic machines} placed in contact with a hot and cold
thermal baths subjected to a non-conservative driving worksource. 
Despite their simplicity, these models showcase an intricate array of phenomena, including pump and heat engine regimes as well as a discontinuous phase transition. %\INM{We will look at two different approaches: Analytically solvable minimal models and those frameworks that have more complex topologies and need to be analyzed numerically.}
We will look at three distinctive topologies: a minimal and beyond minimal (homogeneous and heterogeneous interaction structures). The former case is represented by stark different networks  (``all-to-all" interactions and only a central interacting to its neighbors) and present exact solutions, whereas homogeneous and heterogeneous structures have been analyzed by numerical simulations. 
We find that the topology plays a major role on the thermodynamic performance {for smaller values of individual energies, in part due to the presence of first-order phase-transitions.
%, but that its role becomes negligible if the individual energies are high.
%{If the individual energies are large, the topology is not important and results are well-described by a system with all-to-all interactions.}
%The network of interactions together with the interplay among parameters %(individual and interaction energies, driving and temperature of thermal baths) %may play a fundamental role over the system performance, 
Contrariwise, the topology becomes less important as individual energies  increases and results are well-described by a system with all-to-all interactions.}
%Remarkably, collective effects come from even minimal interaction structures achieving ideal performances provided the number of sites are large. On the other hand, large values of individual energies also ensure system performances but the influence of topology is less important. In such case, results are well described by the all-to-all case in which approximate expressions for thermodynamic quantities are obtained for the thermodynamic limit.

\end{abstract}

\maketitle

\section{Introduction}

The study of thermal engines has always been a central part of thermodynamics \cite{carnot1978reflexions,curzon1975efficiency}. In particular, the last couple of decades have seen a surge of interest in the %\INM{physics of these systems due to the emergency of stochastic thermodynamics}%
thermodynamics of thermal engines due to the emergence of stochastic thermodynamics \cite{seifert2012stochastic,broeck15}. One can for example think about the maximization of power and efficiency \cite{verley2014unlikely, schmiedl2007efficiency,
cleuren2015universality, van2005thermodynamic, esposito2010quantum,
seifert2011efficiency, izumida2012efficiency, 
golubeva2012efficiency, holubec2014exactly, 
bauer2016optimal, 
karel2016, 
tu2008efficiency,
ciliberto2017experiments,bonanca2019,mamede2021obtaining,karel2016prl}; and the influence of studies of system-bath coupling \cite{noa2021efficient,harunari2020maximal} and the level of control \cite{RevModPhys.91.045001,deffner2020thermodynamic,pancotti2020speed,2206.02337}.

One particularly interesting idea is that systems performance might be enhanced by collective effects such as phase-transitions \cite{hooyberghs2013efficiency,campisi2016power}.
These types of collective behavior, e.g., order-disorder phase transitions \cite{yeomans1992statistical} and synchronization \cite{torres2005kuramoto,tonjes2021coherence}, have been observed in a broad range of systems such as complex networks  \cite{bonifazi2009gabaergic,schneidman2006weak,buzsaki2014log,gal2017rich}, biological systems \cite{rapoport1970sodium, gnesotto2018broken,lynn2021broken,smith2019public}  and quantum systems \cite{mukherjee2021many,niedenzu2018cooperative,kurizki2015quantum,lee2022quantum,campisi2016power,mukherjee2020universal,halpern2019quantum,kim2022photonic,PhysRevApplied.19.034023,PhysRevResearch.1.033192,chen2019interaction}.{This has inspired the development of {theoretical models of} thermodynamic engines, in which the performance can be  boosted via collective effects}. Most of these models, however, focus on either one-dimensional systems\cite{mamede2021obtaining,fogedby2017minimal,imparato2021out,imparato2019} or mean-field like models \cite{gatien,vroylandt2020efficiency,herpich,herpich2,forao2023powerful}. Little is known about the influence that network topology might have on system performance.

{
This paper aims to partially fill this gap, by studying the
influence of topology of a simple class of system, composed of interacting units, referred to here as a collection of
nanomachines, placed in contact with a hot and cold ther-
mal baths subjected to a non-conservative driving worksource.
The approach to be considered here is akin to the commonly
referred to as ”lattice-gas” models in the realm of equilibrium
statistical mechanics. They have a longstanding importance
in the context of collective effects and serving as the corner-
stone for numerous theoretical, experimental, and technological breakthroughs, encompassing the ferromagnetism, liquid
phases, the topology effect, the fluctuation-driven generation
of new phases and others, highlighting that distinct systems
have been described/characterized via Hamiltonian of the fundamental models (e.g., the Ising, Potts, XY and Heisenberg
models).
%They have a longstanding importance in the context of collective effects  and are at the heart of numerous theoretical, experimental and technological advances, such as ferromagnetism, liquid phases, the effect of topology, the fluctuation-driven generation of new phases and others, highlighting that distinct systems  have been described/characterized via Hamiltonian of the fundamental magnetism models (e.g., the Ising, Potts, XY and Heisenberg models). 
The all-to-all version for our model}
has been investigated previously \cite{cleuren2001ising,gatien,vroylandt2020efficiency} for finite
and infinite number of interacting units, in which the
cooperative effect gives rise to a
rich behavior, including the enhancement of 
the power and efficiency  at optimal interactions, the existence of
distinct operation models and a discontinuous phase transition. Furthermore, systems with all-to-all interaction can be solved analytically, which makes them easier to analyze. There are, however, also many systems where systems only interact locally (nearest-neighbor like).

In this paper, we present a detailed study on the influence of the topology in {aforementioned
class of interacting units}. 
We will focus on two distinctive approaches: minimal models, which can be treated analytically, and more complicated systems, where we will focus on numerical analysis. In the former class, we will focus on systems with all-to-all interaction and a one-to-all interaction (also known as stargraph), in which a single central spin is interacting with all other units.
After that, we go beyond the minimal models by considering the influence of homogeneous and heterogeneous interaction topologies.
{We will show that, for small values of  individual energies $\beta_\nu\epsilon$ of each occupied unit, the lattice topology can have a significant influence on the system performance, in which the increase of interaction $V$ among units can give rise to a discontinuous phase transition. Conversely, as $\beta_\nu\epsilon$ increases, the phase transition is absent and the topology plays no crucial role and the models seem mutually similar}. In this case one can get approximate expressions for thermodynamic quantities through a phenomenological two-state model.
%[Results show that the lattice topology can have significant influence on the system performance, depending on the engine projection, mainly when individual energies are comparatively smaller than the interaction strength. Conversely, for larger ones, the lattice topology plays no crucial role and the models seem mutually similar. In this case one can get approximate expressions for thermodynamic quantities via a phenomenological two-state model.]

The paper is structured as follows: in Section \ref{Model}, we introduce the model and its thermodynamics.  Section \ref{Topologies}  describes the lattice topologies which will be analysed. In Section \ref{minimal}, the aforementioned minimal models are studied. In Section \ref{Beyond}, the more complicated topologies are studied. Conclusions are drawn in Section \ref{Conclusions}.

\section{Model and Thermodynamics}\label{Model}

{We are assuming a  system composed of $N$ interacting two-state nanomachines. The two states of each individual machine are denoted as $\sigma_i=0(1)$ according to whether it occupies the lower(upper) state  with  energy $0(\epsilon)$. We will consider that the system is in contact with two thermal baths at different temperatures. Furthermore, there will be a non-conservative force (described below) that extracts work from the system, in this way creating a thermal engine.
The state of the full system is then described by $\sigma\equiv(\sigma_1,\sigma_2,...,\sigma_i,...,\sigma_N)$, where $\sigma_i$ describes the state of the $i$'th machine.}
Throughout this paper, we shall restrict  our analysis on  transitions between configurations $\sigma$ and $\sigma^i$
differ by the state of one machine, namely that of unit $i$. In this case,  the time evolution of probability $p(\sigma,t)$ satisfies a  master equation,
\begin{equation} 
\label{eq1} 
    {\dot p}(\sigma,t)=\sum_{\nu=1}^2\sum_{i=1}^N\{\omega_i^{(\nu)}(\sigma^i)p(\sigma^i,t)-\omega_i^{(\nu)}(\sigma)p(\sigma,t) \},
    \end{equation}
where $\sigma^i\equiv(\sigma_1,...,1-\sigma_i,...,\sigma_N)$ and index  $\nu=1(2)$ accounts for transitions induced by the cold (hot) thermal bath. The  transition rate due to
the contact with the $\nu$-th thermal bath
are assumed to be of Arrhenius form  
 \begin{equation}\omega_i^{(\nu)}(\sigma)=\Gamma e^{-\frac{\beta_\nu}{2}\{E_a+{\Delta E}_i(\sigma)+F_\nu\}},
 \label{eq:omegai}\end{equation} 
 where ${\Delta E}_i(\sigma)$ is the difference of energy between states $\sigma^i$ and $\sigma$ { and $\Gamma e^{-\beta_\nu E_a/2}$ accounts to the coupling between the QD and thermal bath, expressed in terms of the activation energy $E_a$ and $\beta_\nu=1/T_\nu$, [hereafter we shall adopt $k_B=1$].  As stated previously,
the interaction among units will depend on the lattice topology,  whose energy of system is given by the generic expression:
\begin{equation}
E(\sigma)=\epsilon n+\sum_{i=1}^N\frac{V}{\langle k\rangle}\sum_{j=1}^{k_i}(\delta_{\sigma_i,1-\sigma_{i+j}}+\delta_{1-\sigma_i,\sigma_{i+j}}),
\label{interac}
\end{equation}
where $n=\sum_{i=1}^N\sigma_i$ denotes the total number of units in the state of energy  $\epsilon$,  $V$ is the interaction strength ($\delta$ being the Kronecker delta) and $\langle k\rangle$ is the average number of neighbours to which each {unit} is connected. Eq.~(\ref{interac}) has been inspired by earlier studies about interacting system,   in which a similar type of interaction
is  consider to describe the interaction between nanomachines in distinct states
\cite{gatien,vroylandt2020efficiency}. Also, this interaction shares some similarities with recent papers \cite{PhysRevResearch.5.023155,cuetara2015double}  in which the tunneling between  two quantum-dots
is investigated via the inclusion of a similar term.
We will both look at topologies where $k_i$,  the number of nearest neighbors of {the unit} $i$ is independent of $i$, $\langle k\rangle=k_i$ (all-to-all interactions and homogeneous systems) and cases where $k_i$ depends on $i$ (stargraph and heterogeneous systems). }
One of these earlier studies also used similar types of work sources:
we consider the worksource  given by $F_\nu$ with $F_\nu=(-1)^\nu(1-2\sigma_i)F$, in such a way that transitions $0\rightarrow 1$ ($1\rightarrow 0$) are favored according to whether the system is placed in contact with the cold (hot) thermal baths.  This type of interaction can also be mapped on other types of systems such as kinesin \cite{liepelt1,liepelt2}, photo-acids \cite{berton2020thermodynamics} and ATP-driven chaperones \cite{de2014hsp70}.
%(consistent to the local detailed balance %$\omega_i^{(\nu)}(\sigma^i)/\omega^{(\nu)}(\sigma)=e^{-\beta_\nu\{{\Delta %E}_i(\sigma)+(-1)^{\nu}d\}}$)\langle E(\sigma)\rangle

From Eq.~(\ref{eq1}) together transition rates given by Eq.~(\ref{eq:omegai}), the time evolution of mean density  $\rho=\langle \sigma_i\rangle$ and mean energy $\langle E(\sigma)\rangle=\sum_{\sigma}E(\sigma) p(\sigma,t)$ obey the following expressions:
\begin{equation}
\label{rho}
{\dot \rho}=\langle(1-2\sigma_i)(\omega^{(1)}_{i}(\sigma)+\omega^{(2)}_{i}(\sigma)) \rangle,
\end{equation}
and
\begin{equation}
\label{en}
\frac{d}{dt}\langle E(\sigma)\rangle={\cal P}+\langle \dot{Q}_1\rangle+\langle\dot{Q}_2\rangle,
\end{equation}
respectively, where ${\cal P}$ and $\langle\dot{Q}_\nu\rangle$ denote the mean
power and the heat exchanged with
the $\nu$-th thermal bath and are given by \cite{gatien,forao2023powerful}:
\begin{equation}
\label{work}
{\cal P}=F\sum_{i=1}^N\left\langle\omega^{(1)}_{i}(\sigma)- \omega^{(2)}_{i}(\sigma) \right\rangle,\\
\end{equation}
and 
\begin{equation}
\label{heat1}
\langle \dot{Q}_\nu\rangle=\langle(\Delta E_i(\sigma)+F(-1)^{(\nu)})\omega_i^{(\nu)}(\sigma) \rangle,
\end{equation}
the standard stochastic thermodynamics expressions for power and heat respectively \cite{seifert2012stochastic}.

Throughout this paper, we will assume that the system has relaxed to a steady state, $p(\sigma,t) \rightarrow p^{st}(\sigma)$, i.e., ${\cal P}+\langle\dot{Q}_1\rangle+\langle\dot{Q}_2\rangle=0$. 
In this case, one can also write the entropy production as
\begin{equation}
 \langle {\dot \sigma} \rangle = \sum_{\nu=1}^2\sum_{\sigma}\sum_{i=1}^N\,{J}^{(\nu)}(\sigma^i)\log\frac{\omega^{(\nu)}_i(\sigma^i) }{\omega^{(\nu)}_i(\sigma)},
\label{eq20}
\end{equation}
with ${J}^{(\nu)}(\sigma^i)=\omega^{(\nu)}_i(\sigma^i)p^{\rm st}(\sigma^i)-\omega^{(\nu)}_i(\sigma)p^{\rm st}(\sigma)$.
%\begin{equation}
%  \langle {\dot \sigma} \rangle=\sum_{\nu=1}^2\sum_{i=1}^N\left\langle %\omega^{(\nu)}_i(\sigma) \log \frac{\omega^{(\nu)}_i (\sigma)}{\omega^{(\nu)}_i %%%(\sigma^{i})}\right\rangle.
%  \label{eqep3}
%\end{equation}
{One can verify from Eq.~\eqref{eq:omegai} that the entropy production, Eq.~(\ref{eq20})
assumes the classical form $\langle {\dot \sigma} \rangle=-\beta_1\langle\dot{Q}_1\rangle-\beta_2\langle\dot Q_2\rangle$, in similarity with equilibrium thermodynamics}.

Under the correct choice of parameters, 
an amount of heat extracted from the hot bath 
$\langle\dot{Q}_2\rangle>0$ can be partially
converted into power output $ {\cal P}<0$ and the system can be used as a heat engine. The efficiency is then defined as $\eta=-{\cal P}/\langle\dot{Q}_2\rangle$, which satisfy the classical relation $\eta\leq\eta_c=1-\beta_2/\beta_1$.

%From the first law of thermodynamics in the steady state,  the entropy production can be re-expressed in terms of $\langle \dot{Q}_2\rangle$ and  ${\cal P}$ as $\langle {\dot \sigma} \rangle = \beta_1{\cal P}+\left(\beta_1 - \beta_2\right) \langle \dot{Q}_2\rangle$, whose ratio between  components is related to the efficiency by means of the relation ${\hat \eta} = \eta/\eta_{\textrm{c}}$, solely differing from  $\eta$ by Carnot bound $\eta_{\textrm{c}}$. 

\section{Lattice Topologies}\label{Topologies}
As stated in the previous section,  we intend to study the differences in thermodynamic
 quantities between
 topologies. We will focus on two classes of systems: Minimal structures, namely stargraph and all-to-all interacting systems, and beyond
minimal structures, comprising homogeneous and heterogeneous systems. 
%Fig. \ref{fig1a} exemplifies their main features and neighborhood %involved.
%\begin{figure}[H]
%    \centering
%    \includegraphics[scale=0.3]{fig0.pdf}
%    \includegraphics[scale=0.18]{rr1.jpeg}
%    \caption{Sketch of lattice topologies along this paper: Star-graphs %$(a)$, regular $(b)$,   random-regular $(c)$ and heterogeneous $(d)$. In %$(a)-(c)$, the central site (red) has $N-1,k$ and $k$, respectively, %whereas
%    in $d$ each neighborhood is generated according to
%    the distribution $P(k)\sim k^{-\gamma}$, where $\gamma=3$.}
%    \label{fig1a}
%\end{figure}
In stargraph systems the interactions are restricted to  a central {unit}(hub) which interacts with its all nearest
neighbor sites (leaves).

Homogeneous and heterogeneous structures present remarkably different properties and has been subject of extensive investigation. While the former case have been largely studied  for addressing the main properties of  graphs, the latter describes a broad class of  systems, such as ecosystems, the Internet, the spreading of rumors and news, citations  and others, in which the agents form heterogeneous networks and are approximately scale-free, containing few nodes (called hubs) with unusually high degree as compared to the other nodes of the network. 
For the homogeneous case, we shall consider those characterized by a fixed neighborhood per unit, being grouped out in two categories,  including a regular arrangement (interaction between nearest neighbors) or  a
random-regular structure, in which all units have the same number of
nearest neighbors, but they are randomly distributed.  Such latter case is commonly generated through a configurational by Bollob\'as \cite{bollobas1980probabilistic}.
%For  $k$ held fixed, a set of $Nk$ points, distributed in $N$ groups, each one containing
%precisely $k$ points is considered. 
%Next,  a random pair of such points is selected and then creates a network
%linking the nodes $i$ and $j$ if there is a pair containing points in the $i$-th and $j$-%th sets until $Nk/2$ pairs (links) are
%obtained. In the case the resulting network configuration presents a loop or duplicate %links, the above process is
%restarted. The latter 
%is a simplified version of the Erd\"os–R\'enyi model \cite{karonski1997origins}.
Finally, among the distinct heterogeneous structures, we will consider the Barabasi–Albert scale-free network, being probably the most well-known example of heterogeneous networks \cite{barabasi1999emergence}.
 The Barabási–Albert (BA) model  is based on a preferential attachment mechanism,
in which the degree distribution follows a power-law with scaling exponent $\gamma=3$ \cite{barabasi1999emergence}.

\section{Minimal models for collective effects: All-to-all interactions versus stargraph}\label{minimal}
 We will first look at the thermodynamic properties of ``all-to-all" and stargraph minimal topologies. There are several reasons for this. First, both of these models can in principle be solved exactly. Second, these structures can be seen as each others opposite. Third, the thermodynamic properties stargraph topologies can give some insights about  heterogeneous networks (e.g. Barabasi-Albert), in a which some nodes are highly connected and most the remaining ones have few connections \cite{barrat2008dynamical,tonjes2021coherence,PhysRevE.92.012904}.

\subsection{Steady-state distribution}
For an all-to-all topology, the state of the system is fully characterized by the number of units {in the upper state}, $n=\sum_{i=1}^{N}\sigma_i$. In terms of total occupation, the master equation for the all-to-all system simplifies to
\begin{equation}
    \centering
    \dot{p}(n,t)=\sum_{\nu=1}^{2}\sum_{\alpha\in\{-1,1\}}[\omega^{(\nu)}_{n,n+\alpha}p(n+\alpha,t)-\omega^{(\nu)}_{n+\alpha,n}p(n,t)],
    \label{solal}
\end{equation}
%in the latter case,
%where for each configuration
%$(n,c)$ one has $(n,c)\rightarrow (n\pm 1, c)$ or $(n,1-c)$

The steady-state distribution for $p^{st}(n)$ then satisfies \cite{gatien}
\begin{equation}
    \centering
    p^{st}(n)=\frac{1}{Z}\left[\prod_{m=0}^{n-1}\omega_{m+1,m}\right]\left[\prod_{m=n+1}^{N}\omega_{m-1,m}\right],
    \label{alltoall}
\end{equation}
where  $Z$ is the normalization factor and $\omega_{ij}=\sum_{\nu=1}^2\omega^{(\nu)}_{ij}$, with transition rates solely expressed in terms of $n$
by $ \omega_{m+1,m}^{(\nu)}=\Gamma (N-m)e^{-\frac{\beta_\nu}{2}\{E_a+\epsilon+{\Delta E}+(-1)^\nu F\}}$ and  $\omega_{m-1,m}^{(\nu)}=\Gamma m e^{-\frac{\beta_\nu}{2}\{E_a-\epsilon- {\Delta E}-(-1)^\nu F\}}$ with $\Delta E=Vm(N-m)/N$.
The thermodynamic quantities can be evaluated from the probability distribution such that,
\begin{eqnarray}
\label{work}
{\cal P} &=&F\sum_{n=0}^{N-1}(J^{(1)}_{n+1,n}-J^{(2)}_{n+1,n}),\\
\langle \dot{Q}_\nu\rangle &=&  \sum_{n=0}^{N-1}\left[\epsilon+\Delta E+(-1)^{\nu} F\right]J^{(\nu)}_{n+1,n},
\label{heat}
\end{eqnarray}
expressed in terms of the probability current $J^{(\nu)}_{n+1,n}=\omega_{n+1,n}^{(\nu)}p^{st}(n)-\omega_{n,n+1}^{(\nu)}p^{st}(n+1)$.
{An overview of the model features in all-to-all topologies will
be depicted next (see e.g. Figs.~\ref{fig2} and Refs.\cite{gatien,vroylandt2020efficiency}), being  strongly dependent on the interplay between individual $\epsilon$
and interaction $V$ parameters.
For $\beta_\nu\epsilon<<1$,  the increase of interaction strength $V$ favors a full occupation
{of units in the upper state} $\rho\rightarrow 1$,  whereas $\rho$
exhibits a  monotonous decreasing behavior upon $V$ is raised for $\beta_\nu\epsilon>>1$. 
The crossover  between above regimes yields for finite $\epsilon$ and depends on $E_a,\beta_1/\beta_2$ and $F$. Another important point to be highlighted concerns that as $E_a$ is increased and $\beta_\nu\epsilon$ is small,  the interaction  marks two distinct trends of the density: its  decreasing behavior of  prior the threshold interaction followed by sharp increase towards $\rho\rightarrow 1$ at $V=V_0$ [see also e.g. Fig.\ref{fig2}$(a)$]. Such
behavior corresponds to a discontinuous phase transition (see e.g. Fig.~\ref{fig5})$(a)$. In Sec.~\ref{secb}, we shall investigate these consequences over the system performance.}

%Although calculable, the generation  of spanning trees for the stargraph is more cumbersome and there is no a simple closed expression like Eq. (\ref{alltoall}) for an arbitrary $N$. Since the number of spanning trees increases faster with the system size than the all-to-all case (e.g. one has $15,56,209,780...$ spanning trees for  $N=3,4,5,6...$), the main expressions for the probability distribution will not be shown here.
It is in principle possible to calculate the exact steady-state distribution for a finite-size star-graph by diagonalising the evolution matrix.
%using the {eigendecomposition of the evolution matrix (as peformed here)} and matrix-tree theorem \cite{schnakenberg1976network}, but the expression becomes more complicated with growing number of leaves.
 {However, some insights into the dynamics can be obtained by doing appropriate approximations, as we will show now. First, we note that the state of the system can be written in terms of $n$ and $c$, denoting the number of leaves and the hub in the upper state, respectively}. The associated probability distribution, $p(n,c,t)$,  satisfies
%\begin{eqnarray}
% \label{eqhh}
 %   {\dot p(c,n)}&=&\omega^{(c)}_{n,n+1}p(c,n+1)+\omega^{(c)}_{n,n-1}p(c,n-1)+\\
%    &+&\omega^{(c,1-c)}_{n,n}p(1-c,n)-(\omega^{(c)}_{n+1,n}+\omega^{(c)}_{n-1,n}+\omega^{(1-c,c)}_{n,n})p(c,n),\nonumber 
%\end{eqnarray}
%    where   $n=0,1,...,N-1$ and transition rates are rewritten as
%\begin{eqnarray}
%%\omega^{(c)}_{n,n}&=&-\omega^{(c)}_{n+1,n}-\omega^{(c)}_{n-1,n}-\omega^{(1-c,c)}_{n,n},\\
%\omega^{(c)}_{n+ 1,n}&=&\Gamma(N-1-n) \sum_{\nu=1}^2e^{-\frac{\beta_\nu}%{2}\left[E_a+\epsilon+V(1-2c)+F(-1)^{\nu}\right]},\\
%\omega^{(c)}_{n-1,n}&=&\Gamma n \sum_{\nu=1}^2e^{-\frac{\beta_\nu}%{2}\left[E_a-\epsilon- V(1-2c)-F(-1)^{\nu}\right]},\\
%\omega^{(1-c,c)}_{n,n}&=&\Gamma \sum_{\nu=1}^2e^{-\frac{\beta_\nu}{2}\left\%{E_a+ (1-2c)[V(N-1-2n)+\epsilon]+F(-1)^{\nu}\right\}}.
%\label{ratesc}
%\end{eqnarray}
\begin{equation}
    \centering
    \dot{p}(c,n,t)=\sum_{\nu=1}^{2}\sum_{\alpha\in\{-1,1\}}\left(\mathcal{J}^{(c,\nu)}_{n,\alpha}(t)+\mathcal{K}^{(n,\nu)}_{c}(t)\right),
    \label{eqhh}
\end{equation}
where 
\begin{equation}
    \centering
\mathcal{J}^{(c,\nu)}_{n,\alpha}(t)\equiv\omega^{(c,\nu)}_{n,n+\alpha}p(c,n+\alpha,t)-\omega^{(c,\nu)}_{n+\alpha,n}p(c,n,t),
\label{eqhh2}
\end{equation}
and
\begin{equation}
    \centering
\mathcal{K}^{(n,\nu)}_{c}(t)=\kappa^{(n,\nu)}_{c,1-c}p(1-c,n,t)-\kappa^{(n,\nu)}_{1-c,c}p(c,n,t),
\label{eqhh3}
\end{equation}
with $n=0,1,...,N-1$ and transition rates rewritten in the following way
\begin{eqnarray}
%\omega^{(c)}_{n,n}&=&-\omega^{(c)}_{n+1,n}-\omega^{(c)}_{n-1,n}-\omega^{(1-c,c)}_{n,n},\\
\omega^{(c,\nu)}_{n+ 1,n}&=&\Gamma(N-1-n) e^{-\frac{\beta_\nu}{2}\left[E_a+\epsilon+V(1-2c)+F(-1)^{\nu}\right]},\\
\omega^{(c,\nu)}_{n-1,n}&=&\Gamma n e^{-\frac{\beta_\nu}{2}\left[E_a-\epsilon- V(1-2c)-F(-1)^{\nu}\right]},\\
\kappa^{(n,\nu)}_{1-c,c}&=&\Gamma e^{-\frac{\beta_\nu}{2}\left\{E_a+ (1-2c)[V(N-1-2n)+\epsilon]+F(-1)^{\nu}\right\}}.
\label{ratesc}
\end{eqnarray}

We assume that the hub dynamics evolves into a  faster time-scales than the relaxation of the surrounding leaves, in such a way that it can be assumed/treated as thermalized at a local leaf transition $n\pm 1$. In other words  transitions
are such that 
%$  \omega^{(0,1)}_{n,n}p(1|n)=\omega^{(1,0)}_{n,n}p(0|n)$
$(\kappa^{(n,1)}_{0,1}+\kappa^{(n,2)}_{0,1})p(1|n)=(\kappa^{(n,1)}_{1,0}+\kappa^{(n,2)}_{1,0})p(0|n)$ and
%[$  %\left(\sum_{\nu=1}^{2}\kappa^{(n,\nu)}_{0,1}\right)p(1|n)=\left(\sum_{\nu=1%}^{2}\kappa^{(n,\nu)}_{1,0}\right)p(0|n)$] 
 hence, the joint probability  $p^{st}(1|n)$ is  given by:
%\begin{equation}
%    p^{st}(1|n)=\frac{\omega^{(1,0)}_{n,n}}%{\omega^{(0,1)}_{n,n}+\omega^{(1,0)}_{n,n}},
%    \label{hubb}
%\end{equation}
\begin{equation}
    p^{st}(1|n)=\frac{\kappa^{(n,1)}_{1,0}+\kappa^{(n,2)}_{1,0}}{\kappa^{(n,1)}_{1,0}+\kappa^{(n,2)}_{1,0}+\kappa^{(n,1)}_{0,1}+\kappa^{(n,2)}_{0,1}},
    \label{hubb}
\end{equation}
where $p^{st}(0|n)=1-p^{st}(1|n)$. By summing Eq. (\ref{eqhh}) over $c$, together the property 
$p(n,c)=p(c|n)p(n)$, one arrives at the following equation for the time evolution of probability $p(n,t)$
%\begin{equation}
%    {\dot p}(n)=\omega_{n,n+1}p(n+1)+\omega_{n,n-1}p(n-1)-%(\omega_{n+1,n}+\omega_{n-1,n})p(n),
%    \label{leaves}
%\end{equation}
%where
%\begin{equation}
%    \omega_{n\pm1, n}=p^{st}(0|n)\omega^{(0)}_{n\pm1,n}+p^{st}(1|n)\omega^{(1)}_{n\pm1,n}.
%\end{equation}
\begin{equation}
    {\dot p}(n,t)=\sum_{\nu=1}^{2}\sum_{\alpha\in\{-1,1\}}\vb{\pi}^{(\nu)}_{n,n+\alpha}p(n+\alpha,t)-\vb{\pi}^{(\nu)}_{n+\alpha,n}p(n,t),
    \label{leaves}
\end{equation}
where
\begin{equation}
    \vb{\pi}^{(\nu)}_{n+\alpha,n}=p^{st}(0|n)\omega^{(0,\nu)}_{n+\alpha,n}+p^{st}(1|n)\omega^{(1,\nu)}_{n+\alpha,n}.
\end{equation}

Since Eq. (\ref{leaves})
is  analogous to Eq. (\ref{solal}) for the all-to-all case, the probability
distribution of leaves $p^{st}(n)$ is given by
% \begin{eqnarray}
%    p^{st}(n)=\frac{1}{Z}\left[\prod_{m=0}^{n-1}\omega_{m+1,m} %\right]\left [\prod_{m=n+1}^{N-1} \omega_{m-1,m}\right],
%    \label{leaf}
%\end{eqnarray}
\begin{equation}
    p^{st}(n)=\frac{1}{Z}\left[\prod_{m=0}^{n-1}\vb{\pi}_{m+1,m} \right]\left [\prod_{m=n+1}^{N-1} \vb{\pi}_{m-1,m}\right],
    \label{leaf}
\end{equation}
\noindent in which $\vb{\pi}_{i,j}\equiv\sum_{\nu=1}^{2}\vb{\pi}^{(\nu)}_{i,j}$ and $Z$ is again the corresponding normalization factor.  Thermodynamic properties can be directly evaluated from Eqs. (\ref{hubb}) and (\ref{leaf}), such as the system density given by $\rho=\sum_{n=0}^{N-1}[n+p^{st}(1|n)]p^{st}(n)/N$, where the probability $p_h$ of hub to be{in the upper state
with individual energy $\epsilon$} reads $p_h=\sum_{n=0}^{N-1}p^{st}(1|n)p^{st}(n)$.
As previously, from the probability distribution, thermodynamic quantities are directly evaluated and given by
\begin{equation}
    \centering
    \mathcal{P}=F\sum_{n=0}^{N-1}\left[\mathcal{L}^{(1)}_{n+1,n}-\mathcal{L}^{(2)}_{n+1,n}+\mathcal{K}^{(n,1)}_{1}-\mathcal{K}^{(n,2)}_{1}\right],
    \label{workstar}
\end{equation}
where $\mathcal{L}^{(\nu)}_{n+1,n}=\vb{\pi}^{(\nu)}_{n+1,n}p^{st}(n)-\vb{\pi}^{(\nu)}_{n,n+1}p^{st}(n+1)$ and $p^{st}(c,n)=p^{st}(c|n)p^{st}(n)$ was considered. Likewise, each  heat component $\mom{\dot{{Q}}_\nu}$
 is given by
%\begin{eqnarray}
%    \centering
%    \mom{\dot{{Q}}_\nu}&=&\sum_{n=0}^{N-1}\sum_c\left[\epsilon+V(1-%c)+(-1)^\nu F\right]\{\omega^{(c,\nu)}_{n+1,n}p^{st}(c,n)\nonumber\\&-&\omega^{(c,\nu)}_{n,n+1}p^{st}%(c,n+1)\}+\sum_{n=0}^{N-1}[V(N-1-2n)+\epsilon]+F(-1)^{\nu})\nonumber%\\&&\{\kappa^{(n,\nu)}_{1,0}p^{st}(0,n)-\kappa^{(n,\nu)}_{0,1}p^{st}%(1,n)\}.
%    \label{heatstar}
%\end{eqnarray}
%\begin{equation}
%    \centering
%    \mom{\dot{{Q}}_\nu}=\sum_{n=0}^{N-1}\left[\Delta E+(-1)^\nu %F\right]\mathcal{L}^{(\nu)}_{n+1,n},
%    \label{heatall}
%\end{equation}
\begin{eqnarray}
    \centering
    \mom{\dot{{Q}}_\nu}&=&\sum_{n=1}^{N-1}\sum_c\left[\epsilon+V(1-2c)+(-1)^\nu F\right]\mathcal{J}^{(c,\nu)}_{n+1,-1}\nonumber\\
    &&\\
    &+&\sum_{n=0}^{N-1}[V(N-1-2n)+\epsilon]+F(-1)^{\nu}]\mathcal{K}^{(n,\nu)}_1,\nonumber
    \label{heatstar}
\end{eqnarray}
where  $\mathcal{J}^{(c,\nu)}_{n+1,-1}$ and $\mathcal{K}^{(n,\nu)}_1$
are evaluated from Eqs.~(\ref{eqhh2}) and (\ref{eqhh3}) in the NESS.

%\INM{\begin{eqnarray}
%    \centering
%    \mom{\dot{{Q}}_\nu}&=&\sum_{n=0}^{N-1}\sum_c\left[\epsilon+V(1-c)+%(-1)^\nu F\right]\mathcal{J}^{(c,\nu)}_{n+1,-1}\nonumber\\
%    &&\\
%    &+&\sum_{n=0}^{N-1}[V(N-1-%2n)+\epsilon]+F(-1)^{\nu}]\mathcal{K}^{(n,\nu)}_1,\nonumber
%    \label{heatstar}
%\end{eqnarray}}

%\INM{\begin{align}
%\mathcal{J}^{(c,\nu)}_{n+1,\alpha}&=\omega^{(c,\nu)}_{n+1,n+1+\alpha}p(c,n%+1+\alpha)-\omega_{n+1+\alpha,n+1}p(c,n+1)\\
%\mathcal{J}^{(c,\nu)}_{n+1,-1}&=\omega^{(c,\nu)}_{n+1,n}p(c,n)-%\omega_{n,n+1}p(c,n+1)
%\end{align}}

\begin{figure}
\centering
                        \includegraphics[scale=0.7]{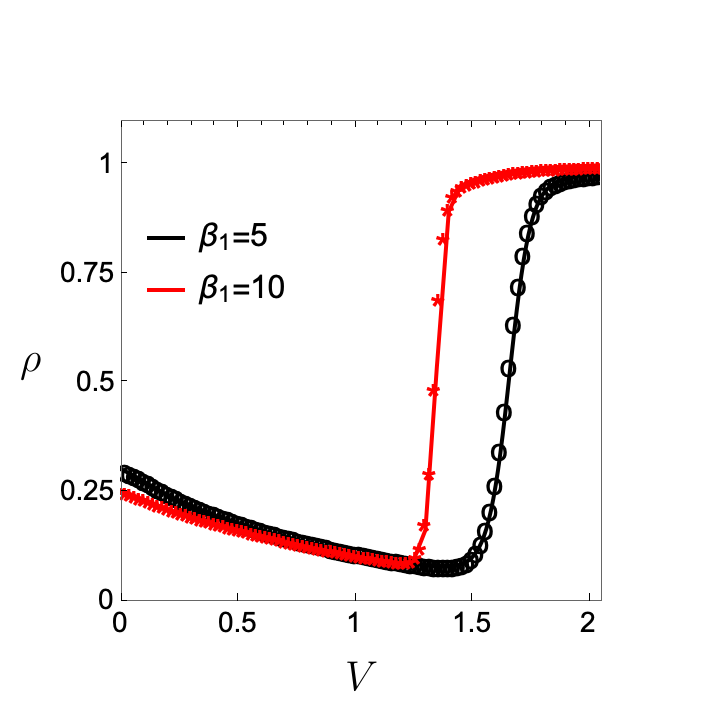} 
               \caption{For the stargraph, the comparison between the exact
               system density $\rho$, obtained by diagonalising the evolution matrix (continuous lines), %{from the eigendecomposition of the evolution matrix} 
               and the approximation Eqs. (\ref{hubb}) and (\ref{leaf}) (symbols). 
               Circles and stars correspond to $\beta_1=5$  and $\beta_1=10$,
               respectively.
              Parameters: $N=30,F=1,E_a=2$ and $\beta_2=1$. 
               }
        \label{comp}
    \end{figure}

Fig. \ref{comp} compares the evaluation of system density $\rho$ from the
exact (continuous lines) method with the approximate (symbols) method. Both curves agree remarkably well.

\subsection{General features and heat maps for finite $N$}\label{secb}
\begin{figure}
                \includegraphics[width=.56\textwidth]{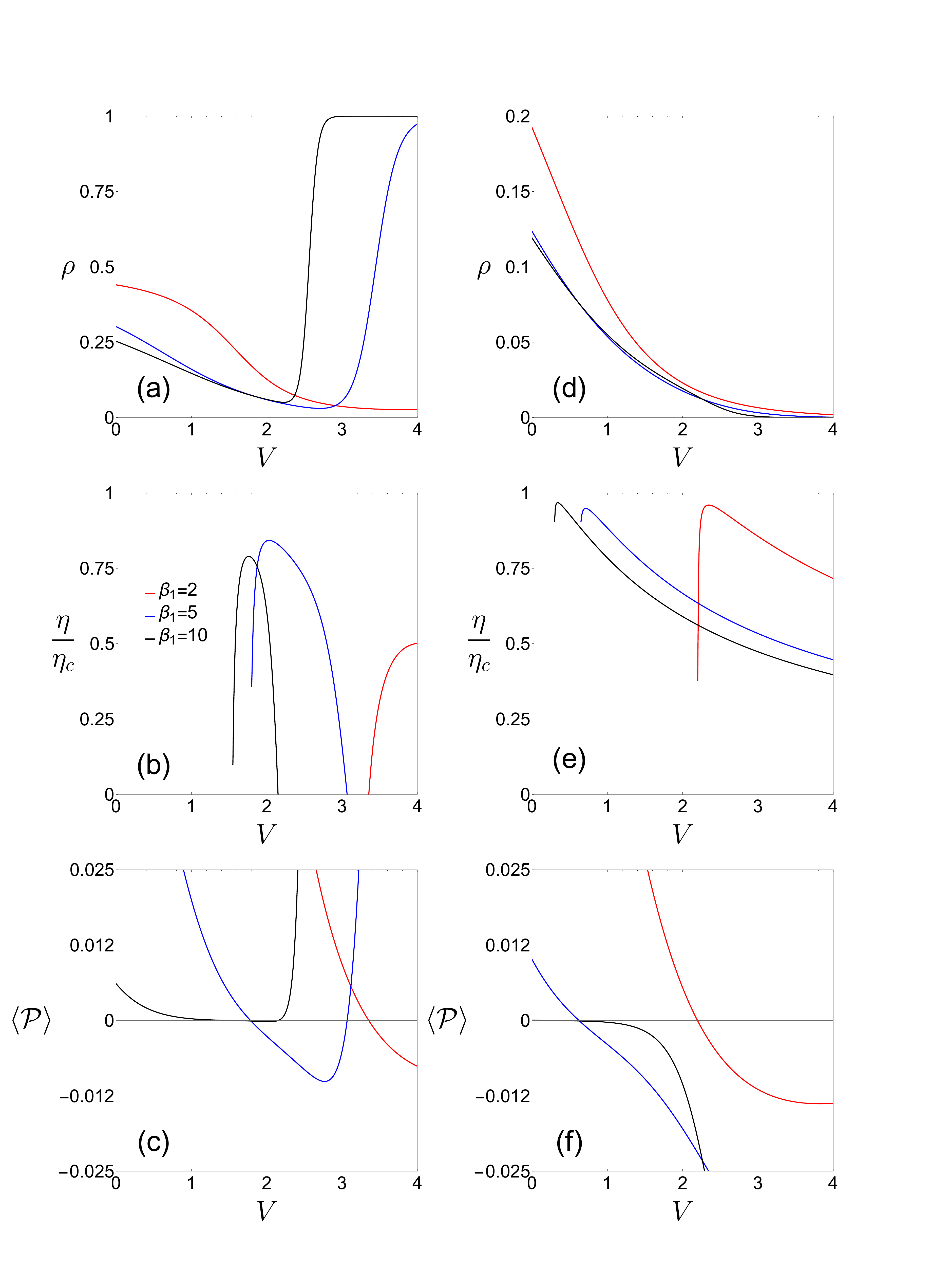} 
               \caption{System density $\rho$ (top), 
               efficiency $\eta/\eta_c$ (center) and power $\langle {\cal P}\rangle\equiv {\cal P}/N$
               per particle (bottom) versus $V$ for the all-to-all case, for $\epsilon=0.1$ (left panels) and $\epsilon=1$ (right panels).  Discontinuities in the efficiency correspond to crossovers from pump-dud and dud-engine regimes. Parameters: 
               $F=1,E_a=2$ and $\beta_2=1$. 
               }
        \label{fig2}
    \end{figure}

\begin{figure}
        \includegraphics[width=.56\textwidth]{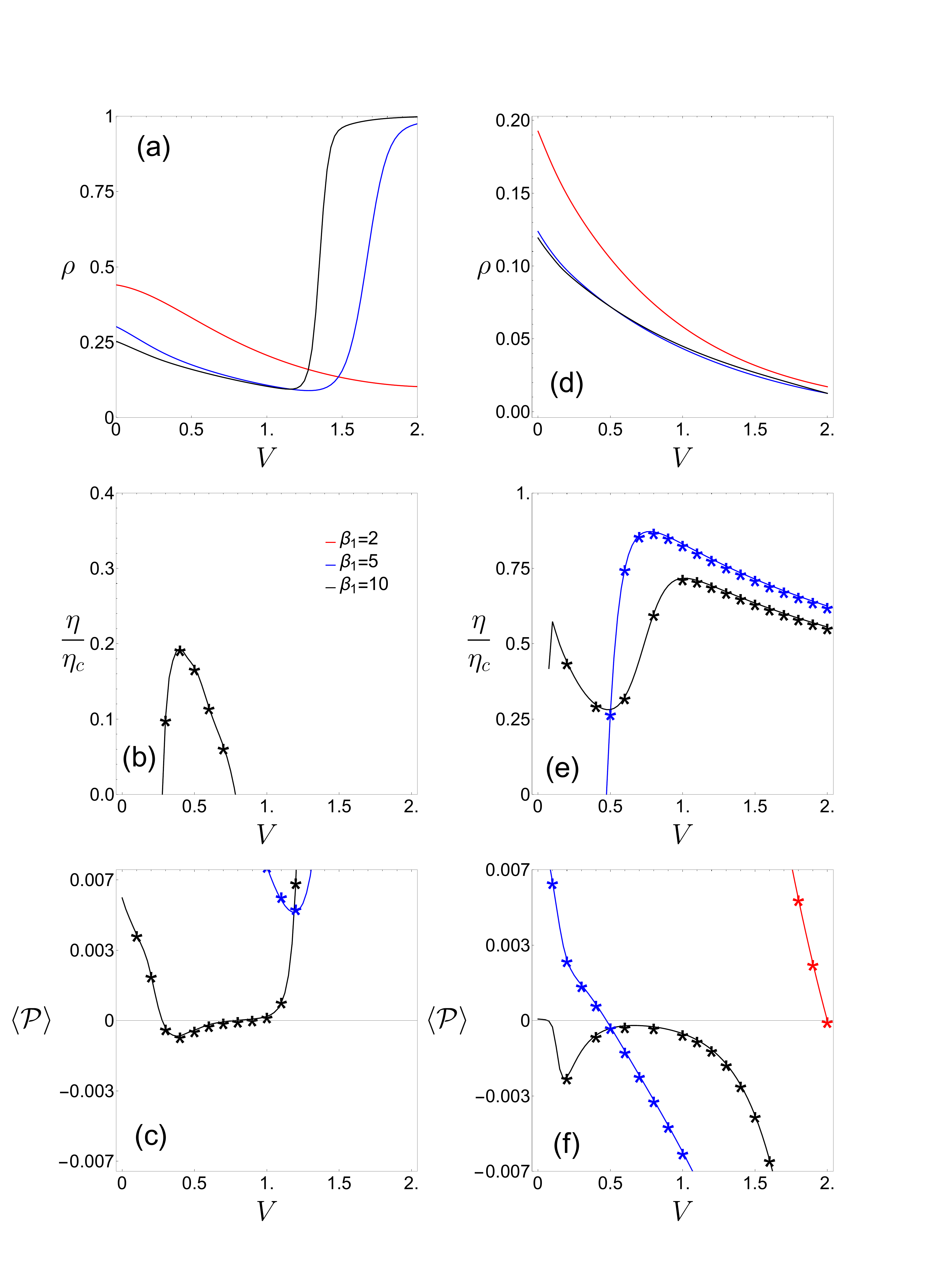}
               \caption{System density $\rho$ (top), 
               efficiency $\eta/\eta_c$ (center) and power $\langle {\cal P}\rangle\equiv {\cal P}/N$
               per particle (bottom) versus $V$ for the stargraph case, for $\epsilon=0.1$ (left panels) and $\epsilon=1$ (right panels). Symbols denote results from numerical simulations. Parameters: 
               $F=1,E_a=2$ and $\beta_2=1$. }
        \label{fig32}
    \end{figure}

To reduce the number of parameters, we will assume that $\beta_2=1$ unless specified otherwise. Furthermore, we will look at $\epsilon=0.1$ and $1$ and vary $\beta_1$.
Figs. \ref{fig2} and \ref{fig32} summarize the main findings about  minimal models for interacting  for a small  system of size $N=20$.
            
{As the all-to-all, the increase of interaction strength $V$ also changes  $\rho$ significantly for the stargraph  and, consequently, affects the engine performance. While intermediate densities favor the system operation as a pump, their  emptying  when $V$ is increased changes the operation regime, from a pump to a heat engine and also increases the engine performance, whose performances are meaningfully different for $\epsilon=0.1$ (smaller) and $1$ (larger) individual  $\beta_\nu\epsilon$'s. }The maximal reachable efficiency $\eta_{ME}$ is always lower than $\eta_c$ for finite $N$, as  expected.

Another common behavior in both cases is the fact that large $V$ favors a full occupation
{of the upper state} when $\beta_\nu\epsilon$ is small (see e.g. panels $(a)$ in Figs. \ref{fig2} and \ref{fig32}), implying that the system operates dudly when most of units are  {in the upper state}, whose  crossover from heat to  dud regime is marked by
a discontinuous phase transition. Conversely,  {the increase of $\epsilon$} marks  the absence
of phase transition for a broader range of  $V$ and consequently not only extends the engine regime
but also improves system performance.     
Despite closely dependent on parameters, both $\eta$ and ${\cal P}$ exhibit  similar trends as $\beta_1$ is raised for the all to all case.
Although having inferior performance than the all-to-all (at least
for the chosen set of parameters), the stargraph yield some striking features for smaller $N$ (see e.g. in Figs.~\ref{fig32}, \ref{fig42} and \ref{fig5}), including an intermediate
sets of $V$ in which both $\eta$ and ${\cal P}$ do not behave monotonously, characterized by a local and global maximum ($\eta_{MP}$) and minimum (${\cal P}_{mP}$), as
can be seen  in Figs. \ref{fig32} $(e)-(f)$. In all cases, exact results (continuous) agree very well with numerical simulations (symbols).

\begin{figure*}
    \centering
    \includegraphics[scale=0.4]{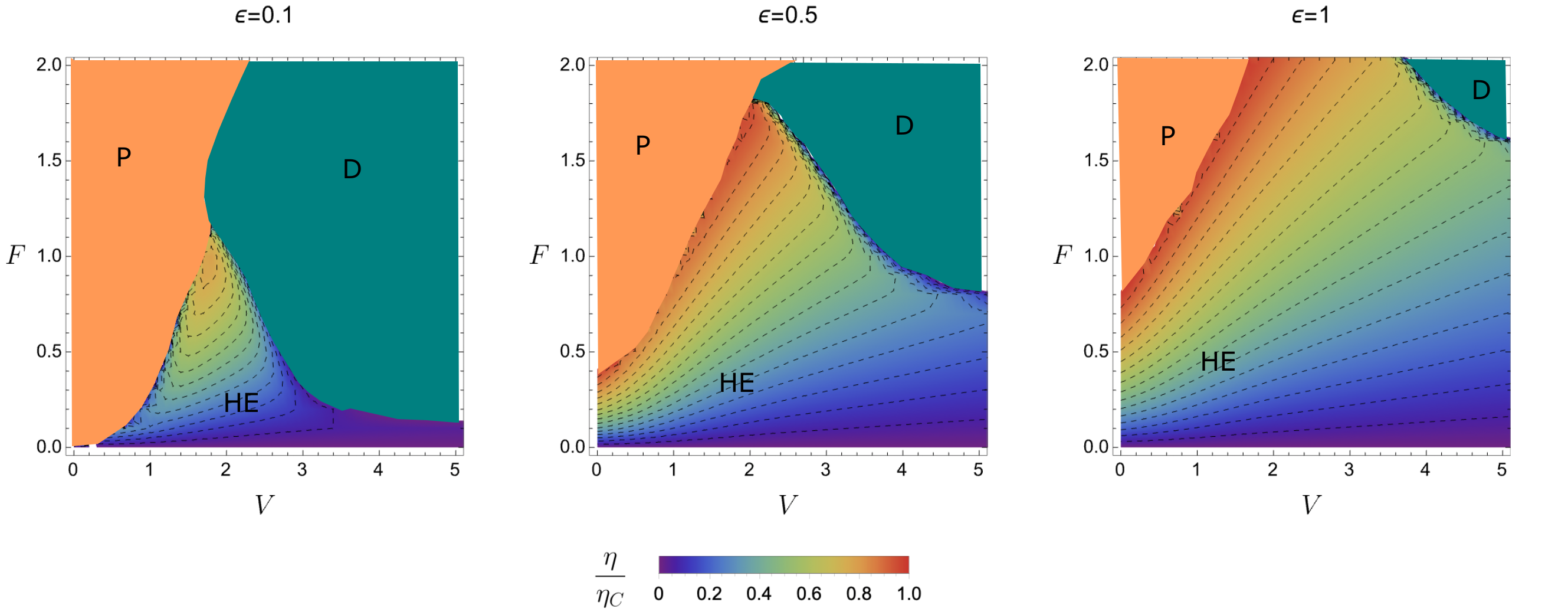}
    \caption{For the all-to-all topology, the efficiency heat maps for various choices of $\epsilon$' as  in Fig. \ref{fig32}. HE, P and D denote the heat engine, pump and dud regimes, respectively. Parameters: $N=20$, $E_a=2$, $\beta_1=10$, $\beta_2=1$.}
    \label{fig41}
\end{figure*}

\begin{figure*}
    \centering
    \includegraphics[scale=0.4]{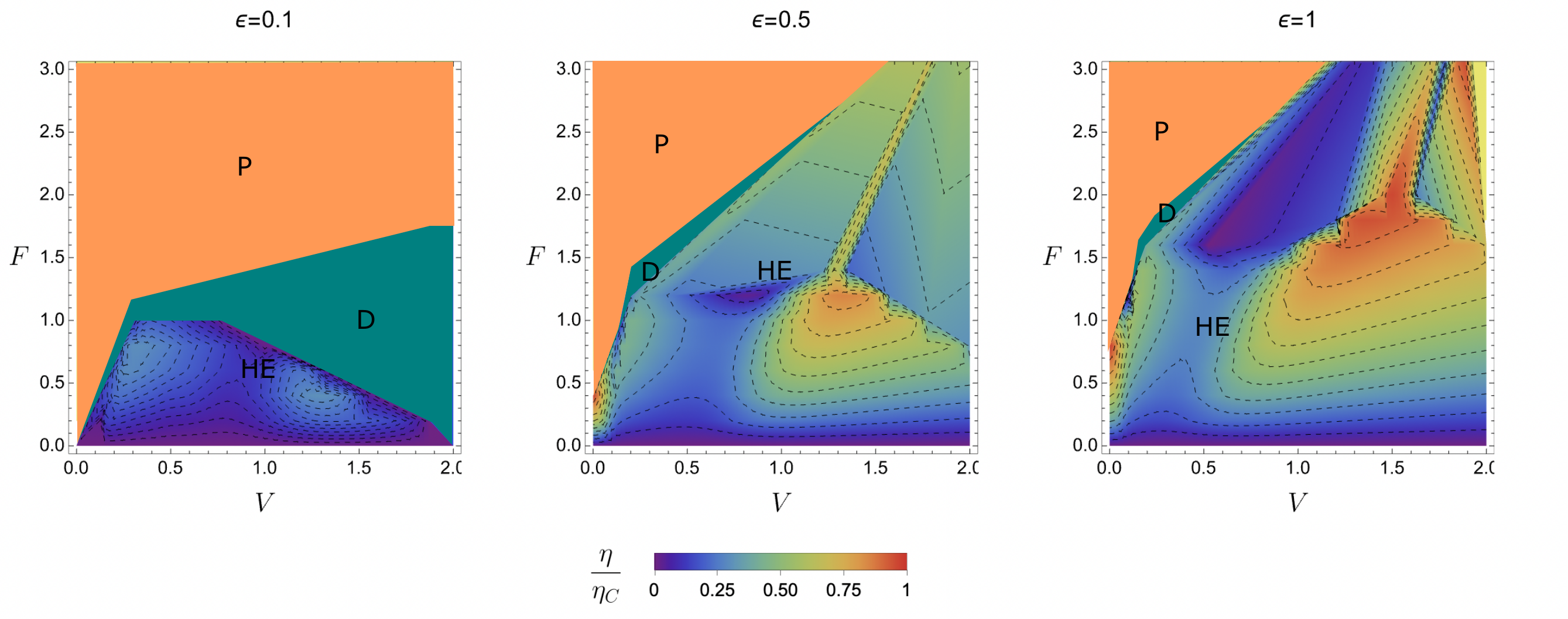}
    \caption{For the stargraph topology, the efficiency heat maps for various choices of $\epsilon$, as in Fig. \ref{fig32}. Parameters: $N=20$, $E_a=2$, $\beta_1=10$, $\beta_2=1$.}
    \label{fig42}
\end{figure*}

A  global phase-portrait is depicted in
Figs. \ref{fig41} and \ref{fig42} for $N=20$. These results are in agreement with the aforementioned  and
reinforce previous findings, including
larger maximum efficiencies and power for  all-to-all interactions than
stargraph ones for small $N$'s, but such later one presents two distinct regions
(for lower and larger $V$'s) which the heat engine
operates more efficiently. Similar findings are shown
in Sec. \ref{app} for ${\cal P}$'s.

{In Secs.~\ref{ivc} and \ref{ivd}, remarkable aspects about both minimal structures, including the existence of a discontinuous transition for
smaller individual energies as well as its suppression as $\epsilon$ increases, shall be described.}

\subsection{Effect of system sizes and phase transitions} \label{ivc}
\begin{figure}
                        \includegraphics[width=.56\textwidth]{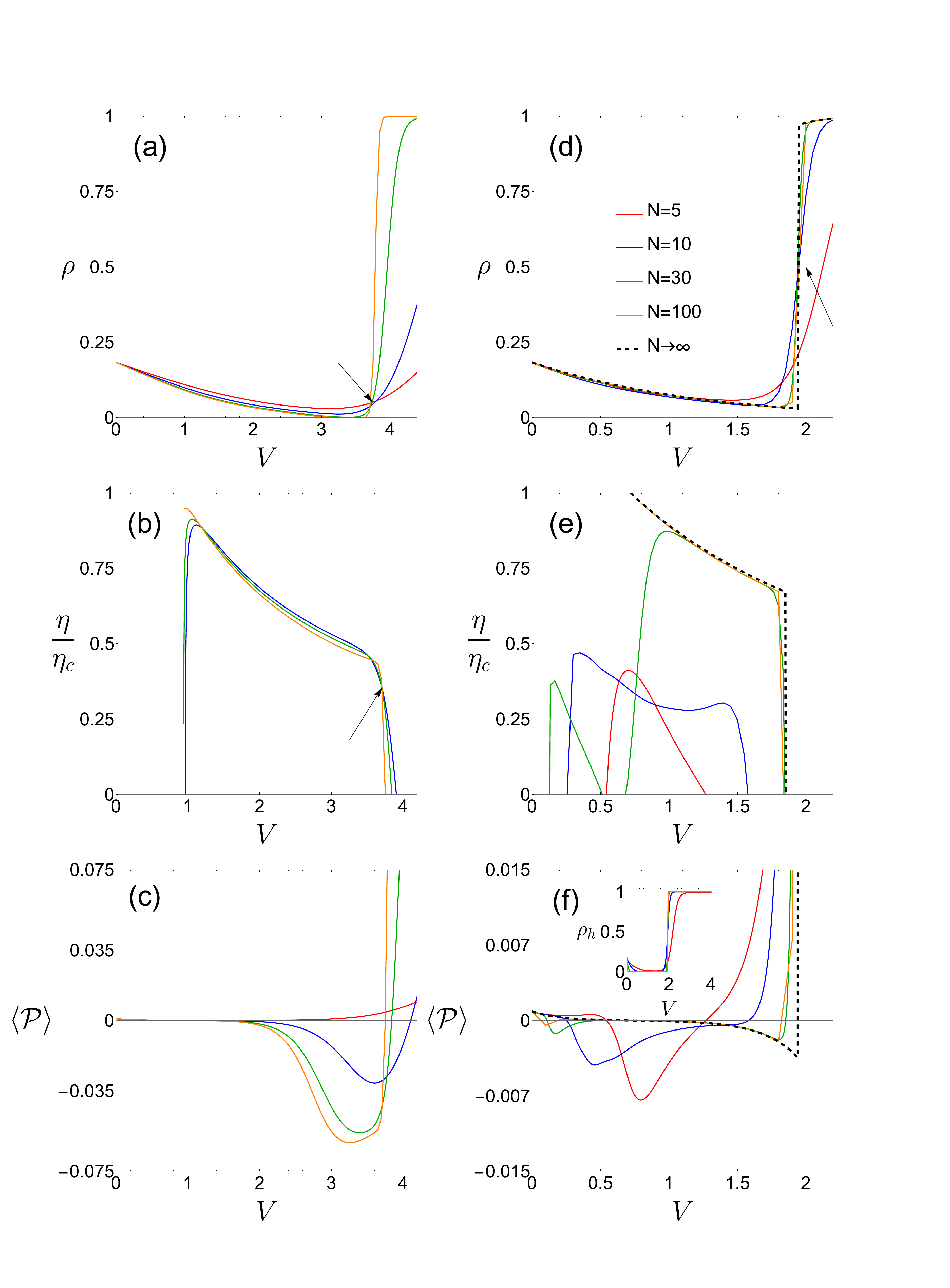}
               \caption{The effect of system size in minimal collectively models: Left and right panels depict the behavior of density (top), $\eta/\eta_c$ (center) and $\langle {\cal P}\rangle\equiv {\cal P}/N$ (bottom) for the all-to-all
               and the stargraph, respectively. Arrows indicate the discontinuous phase transitions, characterized by the crossing among curves. The inset in the right bottom panel indicates the hub density for the stargraph model. Dashed lines: Results for $N\rightarrow \infty$. Parameters:  $F=1,E_a=2,\epsilon=0.5,\beta_1=10$ and $\beta_2=1$. }
        \label{fig5}
    \end{figure}

{The first common aspect regarding the behavior of  stargraph and all-to-all interaction structures
is that the increase
of interaction $V$ (for smaller values of $\beta_\nu\epsilon$) not only influences the system properties and the engine's performance but also
gives rise to a phase transition characterized by a full occupancy of  units
{in the upper state} as $N\rightarrow \infty$. }However, the behavior
of finite systems provides some clues about the classification 
of phase transition, as described by the finite size scaling theory
\cite{odor2008universality,henkel2008non,fioreprl2011,fiore2011comparing,fiorepre2007,fiorefss,fiorefss2}. In the present case, the existence of a crossing among curves for distinct (finite) system
sizes  $N$ reveals a discontinuous phase transition \cite{fiorefss,fiorefss2}, as depicted in
Fig. \ref{fig5}.
% Although similar features are present for $\epsilon=0.1$, the engine performance is superior for $\epsilon=0.5$.
More specifically, the density curves $\rho$ strongly depend on the system size near  phase-transition $V_0$ ($V_0=3.712(3)$ and $1.942(2)$ for the all-to-all and stargraph, respectively), whose  intersection among curves
is consistent to a  density jump for $  N\rightarrow \infty$. Such features
are also
manifested in the behavior of both $\eta$ and ${\cal P}$ (see arrows in Figs. \ref{fig5}), marking the coexistence between  heat engine and dud regimes.
%is a hallmark of discontinuous phase transitions in 
% equilibrium \cite{fioreprl2011,fiore2011comparing,fiorepre2007} 
%and nonequilibrium \cite{fiorefss,fiorefss2,noa2019entropy} systems and %consistent to a  density jump for $  N\rightarrow \infty$, also
%manifested in the behavior of both $\eta$ and ${\cal P}$ (see arrows in Figs. \ref{fig5}), marking the coexistence between  heat engine and dud regimes. The engine performance is also influenced by the increase of the system size due to the enhancement of collective effects, mainly for the stargraph (see e.g. bottom panels in Fig. \ref{fig5}).
{The opposite scenario is verified by raising $\epsilon$, as depicted in Fig.~\ref{figf} for $\epsilon=1$ for both all-to-all and stargraph topologies. Unlike the behavior of $\epsilon=0.5$, the phase transition is absent for both structures and as a consequence, the heat engine regime
is broader.}

We close this section by stressing that, although discontinuous phase transition have already been reported for similar systems \cite{gatien}, the existence of a phase transition  in the stargraph structure is revealing and suggests that a minimal interaction structure is sufficient
for introducing collective effects that are responsible for the phase transition.

\subsection{The $N\rightarrow \infty$ limit and phenomenological descriptions  }\label{ivd}
\begin{figure}
    \centering
    \includegraphics[width=.56\textwidth]{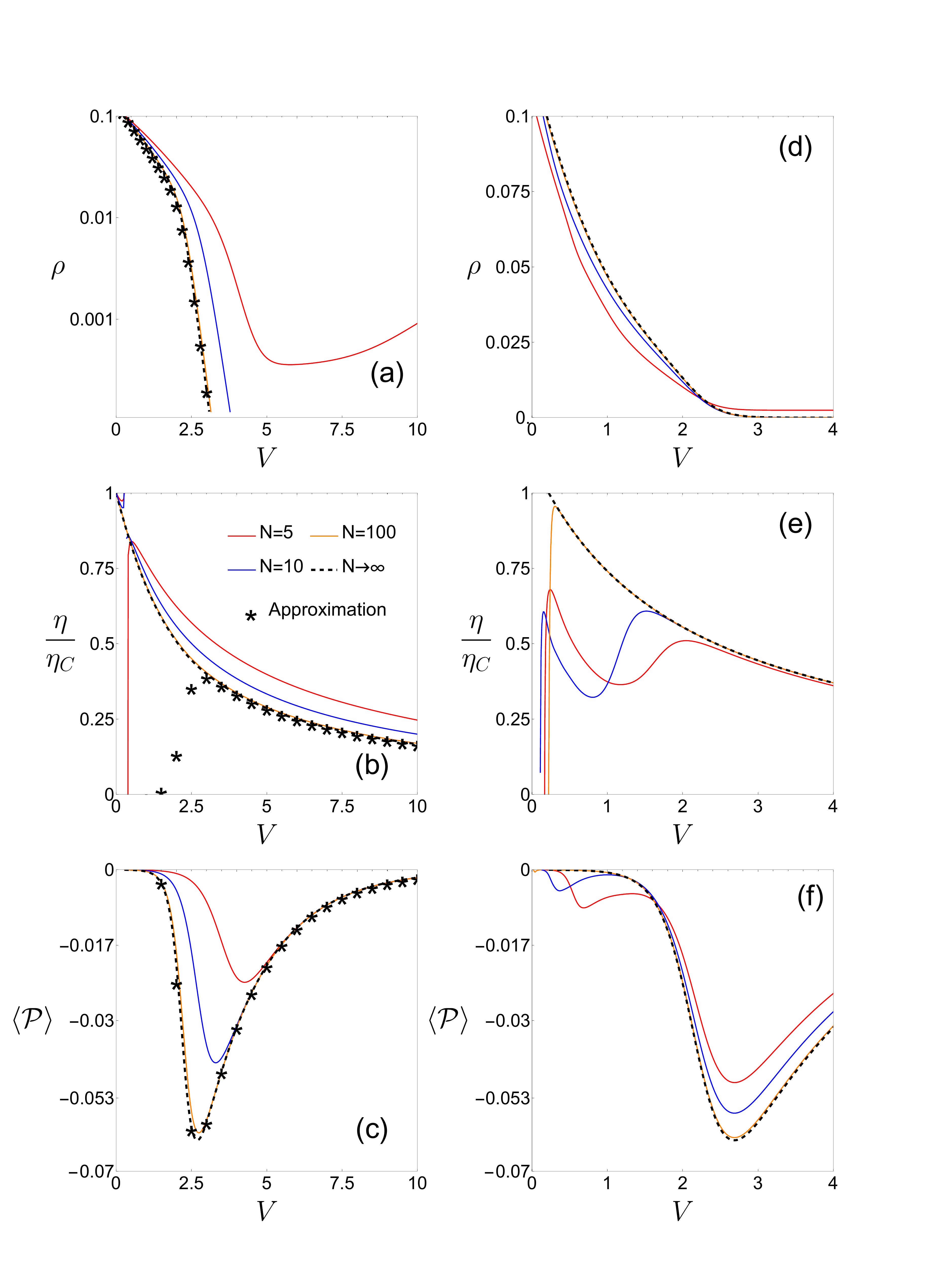}
    \caption{Depiction of system density (top), $\eta/\eta_c$ (center) and $\langle {\cal P}\rangle\equiv {\cal P}/N$ (bottom) for distinct system sizes and also for the $N\rightarrow\infty$ limit for all-to-all(left) and stargraph(right). The monolog plot of
    $\rho$ in $(a)$ has been considered in order to validate Eq.(\ref{theor}).
    Dotted lines: the phenomenological description for the all-to-all case. Parameters: 
$\beta_1=10$, $\beta_2=1$, $E_a=2$, $\epsilon=1$ and $F=1$.}
    \label{figf}
\end{figure}

A question which naturally arises concerns the system behavior in the thermodynamic limit  $ N\rightarrow \infty$
for both all-to-all and stargraph structures. The former case is rather simple and can be derived directly from
transition rates, in which system behavior is described by a master equation with non linear transition rates. Since the all-to-all  dynamics 
is fully characterized {by  the quantity} $n$, 
the macroscopic dynamics is given by
the probability of occupation $p_1$, corresponding to $\rho$ in the thermodynamic limit
$p_1={\lim}_{N\rightarrow \infty}\sum_{i=1}^{N}ip^{st}(i)/N$  ($p_0=1-p_1$) \cite{gatien,forao2023powerful}
 and described by the master equation that has the form $ {\dot p}_1=\sum_{\nu=1}^2J^{(\nu)}_{10}$, such that:
 \begin{equation}
 {\dot p}_1=\sum_{\nu=1}^2[\omega^{(\nu)}_{10}(1-p_{1})-\omega^{(\nu)}_{01}p_1],
 \label{nolinear}
 \end{equation}  
 where  transition rates $\omega^{(\nu)}_{10}$ and  $\omega^{(\nu)}_{01}$ denote the
 transition {the lower to the higher state}and vice versa, respectively, and are listed below:
\begin{eqnarray}
\omega^{(1)}_{10}=\Gamma e^{-\frac{\beta_1}{2}\{E_a+\Delta E_{10} - F\}},\nonumber\\
\omega^{(2)}_{01}=\Gamma e^{-\frac{\beta_2}{2}\{E_a-\Delta E_{10}+ F\}}, 
\label{tra}
\end{eqnarray}
where $\Delta E_{10}=V(1-2p_1)+\epsilon$. 
For $N\rightarrow \infty$,  expressions for the power $\langle{\cal P}\rangle\equiv{\cal P}/N$ and heat per unit $\langle\dot{\mathcal{Q}}_\nu\rangle\equiv\langle\dot{{Q}}_\nu\rangle/N$ from Eqs. (\ref{work})
and (\ref{heat}) read
\begin{eqnarray}
\label{work1}
\langle {\cal P}\rangle=F(J_{10}^{(1)}-J_{10}^{(2)}) \quad {\rm and } \quad\\
\langle \dot{\mathcal{Q}}_\nu\rangle=\left( \Delta E_{10}+(-1)^\nu F\right)J_{10}^{(\nu)}  \label{heati}
\end{eqnarray}
respectively. 
{
 $p_1$ is obtained
by solving Eq.~(\ref{nolinear}). As shown in Sec.~\ref{ivc},
small and large values of individual energy $\beta_\nu\epsilon$ mark different behaviors a $N$ increases, the former yielding
a discontinuous phase transition. Unlike the behavior of finite $N$, the discontinuous phase transition 
is featured by the existence of a hysteretic branch in which the system has}
a bistable behavior \cite{gatien}. {We shall focus on $\epsilon=1$ which describes the behavior
of large $\beta_\nu\epsilon$'s, as depicted in Fig. \ref{figf}, together with a comparison with different $N$'s.  
As can be seen in this figure, in both cases, }maximum efficiencies $\eta_{ME}$'s (for coupling strength V=$V_{ME}$'s) and (absolute) minimum powers ${\cal P}_{mP}$'s (for coupling strength $V=V_{mP}$'s) increase as $N$ is
raised and approaching to the $N\rightarrow \infty$, consistent with enhancing collective effects. However, contrasting with the power, the efficiency
for smaller system sizes is larger for $V>V_{ME}$. This can be understood from the interplay between power and $\langle \dot{Q}_2\rangle$.
For $V>V_{mP}$, the power mildly changes with $N$, whereas $\langle \dot{Q}_2\rangle/N$ monotonically increases with $N$.  Likewise for ${V_{ME}}<V<V_{mP}$, but in this case $\langle \dot{Q}_2\rangle/N$ increases
``faster" than $\langle {\cal P}\rangle$.

Although Eq.~(\ref{nolinear}) can be solved numerically for any set of parameters,  
its nonlinear shape makes  it impossible to obtain analytical results. However, it is possible to get some insights about the system in the heat engine
regime when $p_1<<1$ (and $p_0$ is close to $1$). In this case, {the terms $p_1$ and $p_0$
 appearing in transition rates can be neglected and treated as $p_0\approx 1$, respectively, in such a way one arrives at the following}formula:
\begin{equation}
p_1 \approx \frac{\omega^{(1)}_{10}+\omega^{(2)}_{10}}{\omega^{(1)}_{01}+\omega^{(2)}_{01}+\omega^{(1)}_{10}+\omega^{(2)}_{10}}.
\label{theor}
\end{equation}
By inserting transition rates from Eq. (\ref{tra}) into Eq. (\ref{theor}),
we arrive at the following approximate expression for $p_1$:
\begin{equation}
    \centering
    p_{1}\approx \frac{\mathcal{A}_1 e^{-\frac{1}{2} \beta _1 \left(V+\epsilon-F\right)}+\mathcal{A}_2 e^{-\frac{1}{2} \beta _2 \left(V+\epsilon+F\right)}}{\mathcal{A}_1 e^{\frac{1}{2}
   \beta _1 \left(V+\epsilon-F\right)}+\mathcal{A}_2 e^{\frac{1}{2} \beta _2 \left(V+\epsilon+F\right)}},
   \label{p1f}
\end{equation}
%\begin{equation}
%    \centering
%    p_{1}\approx \frac{\mathcal{A}_1 e^{-\frac{1}{2} \beta _1 \left(\theta %_2+V\right)}+\mathcal{A}_2 e^{-\frac{1}{2} \beta _2 \left(\theta _1+V\right)}}
%{\mathcal{A}_1 e^{\frac{1}{2}
%   \beta _1 \left(\theta _2+V\right)}+\mathcal{A}_2 e^{\frac{1}{2} \beta _2 %\left(\theta _1+V\right)}},
%\end{equation}
where $\mathcal{A}_\nu=e^{-\frac{\beta_\nu}{2}E_a}$.
Approximate expressions for ${\cal P}$ and $\langle Q_\nu\rangle$'s in the heat engine
are promptly obtained inserting Eq. (\ref{p1f}) into Eqs.~(\ref{work1}) and (\ref{heati}), respectively. Although they are cumbersome, they solely depend on the model parameters $\beta_1,\beta_2,E_a,\epsilon,F$ and $V$.
The comparison between exact and approximate results
is also shown in top panels from
Fig. \ref{figf} (symbols) for $\epsilon=1$, in which
no phase transition yields (at least for limited $V$'s).   As can be seen,
the agreement is very good for $p_1<<1$.

The limit $N\rightarrow \infty$ for the stargraph is obtained in a similar way, but leaves
and hub are treated separately. Given that Eqs. (\ref{solal}) and (\ref{leaf}) present similar forms, the density of leaves $p_1={\lim}_{N\rightarrow \infty}\sum_{i=1}^{N-1}ip^{st}(i)/N$  also has
the form of Eq.~(\ref{theor}) and are given by
\begin{equation}
p_1=\frac{\omega^{(c,1)}_{10}+\omega^{(c,2)}_{10}}{\omega^{(c,1)}_{10}+\omega^{(c,1)}_{01}+\omega^{(c,2)}_{10}+\omega^{(c,2)}_{01}},
\label{p1s}
\end{equation}
where  transition rates given by
\begin{eqnarray}
%\omega^{(c)}_{n,n}&=&-\omega^{(c)}_{n+1,n}-\omega^{(c)}_{n-1,n}-\omega^{(1-c,c)}_{n,n},\\
\omega^{(c,\nu)}_{10}&=&\Gamma e^{-\frac{\beta_\nu}{2}\left[E_a+\epsilon+V(1-2c)+F(-1)^{\nu}\right]},\\
\omega^{(c,\nu)}_{01}&=&\Gamma  e^{-\frac{\beta_\nu}{2}\left[E_a-\epsilon- V(1-2c)-F(-1)^{\nu}\right]}.
\label{ratesc}
\end{eqnarray}

Since $p_1$ is dependent on the hub occupation, it is worth investigating
 its behavior when $N\rightarrow \infty$. From 
 Eq.~(\ref{hubb}),  $p^{st}(1|n)\rightarrow 0$ and $1$ for $n<N/2$ and $n>N/2$, respectively, as $N\rightarrow \infty$. Also, $p_h=\sum_{n=0}^{N-1}p^{st}(1|n)p^{st}(n)/N \rightarrow 0$ and $1$ when $p_1$ is small and large, respectively. 
From  Eqs. (\ref{workstar})
and (\ref{heatstar}),  expressions for the power $\langle{\cal P}\rangle\equiv{\cal P}/N$ and heat $\langle\dot{\mathcal{Q}}_\nu\rangle\equiv\langle\dot{{Q}}_\nu\rangle/N$ are obtained 
 by noting that the hub contribution vanishes as limit $N\rightarrow \infty$ and hence they read
\begin{equation}
\langle{\cal P}\rangle=F\left[(\pi_{10}^{(1)}-\pi_{10}^{(2)})(1-p_{1})-(\pi_{01}^{(1)}-\pi_{01}^{(2)})p_{1} \right],
\label{powers}
\end{equation}
and 
\begin{eqnarray}
    \centering
    \mom{\dot{\mathcal{Q}}_\nu}&=&\left[\epsilon+V(1-2c)+(-1)^\nu F\right]\left[\pi_{10}^{(\nu)}-(\pi_{10}^{(\nu)}+\pi_{01}^{(\nu)})p_{1} \right],
    \end{eqnarray}
respectively, where $\pi_{10}^{(\nu)}=\omega^{(0,\nu)}_{10}$ and $\pi_{10}^{(\nu)}=\omega^{(1,\nu)}_{10}$ provided $p_h=0$ and $1$, respectively. Efficiency $\eta$ is straightforwardly evaluated using the above equations
\begin{equation}
\eta=-\frac{F\left[(\pi_{10}^{(1)}-\pi_{10}^{(2)})(1-p_{1})-(\pi_{01}^{(1)}-\pi_{01}^{(2)})p_{1} \right]}{\left[\epsilon+V(1-2c)+ F\right]\left[\pi_{10}^{(2)}-(\pi_{10}^{(2)}+\pi_{01}^{(2)})p_{1} \right]}.
\label{effs}
\end{equation}

We pause again to make a few comments about Eq.~(\ref{effs}). 
For small values of $\beta_\nu\epsilon$, in which a  discontinuous phase transition
yields at $V_0$,   $p_h$ jumps
from $0$ to $1$ 
for $V<V_{0-}$ and $V>V_{0+}$, respectively, and  $p_h=1/2$ precisely at $V=V_0$. Second, from the hub behavior,  it follows that  $p_1$ jumps from
$p_{1-}$ to $p_{1+}$, where 
$p_{1-}$($p_{1+}$) are obtained from Eq.~(\ref{p1s}) evaluated at $c=0$ (for $V\rightarrow V_{0-}$)
and $c=1$ (for $V\rightarrow V_{0+}$), respectively. Third,  
the order-parameter jump is also followed by  discontinuities  in
the behavior of thermodynamic quantities, such as $\langle{\cal P}\rangle$ and $\eta$. They are  evaluated from Eqs.~(\ref{powers}) and (\ref{effs}) at $c=0$ and  $V=V_{0-}$
to $c=1$ and  $V=V_{0+}$, respectively.
Fourth and last, large $\beta_\nu\epsilon$'s mark no phase transitions for limited values of $V$ and hence  the power and efficiency are evaluated at $c=0$ (since $p_h=0$).  All above findings, together the reliability of Eqs.~(\ref{p1s}), (\ref{powers}) and (\ref{effs}) for small  and large values
of $\beta_\nu\epsilon$, are depicted in
bottom panels from Figs.~\ref{fig5} and \ref{figf} {for $\epsilon=0.5$ and $1$, respectively}. As for
the all-to-all, thermodynamics quantities for finite $N$ approach 
to the $N\rightarrow \infty$ limit as $N$ is increased.

 We close this section by 
drawing a comparison between all-to-all and stargraph performances
in the heat engine regime for the same set of parameters in Figs.~\ref{fig5}
and \ref{figf}.
While the former structure is more efficient,  stargraph ones present  larger power outputs.

\section{Beyond the minimal models: Homogeneous and Heterogeneous topologies}\label{Beyond}
\begin{figure}
            \includegraphics[width=.56\textwidth]{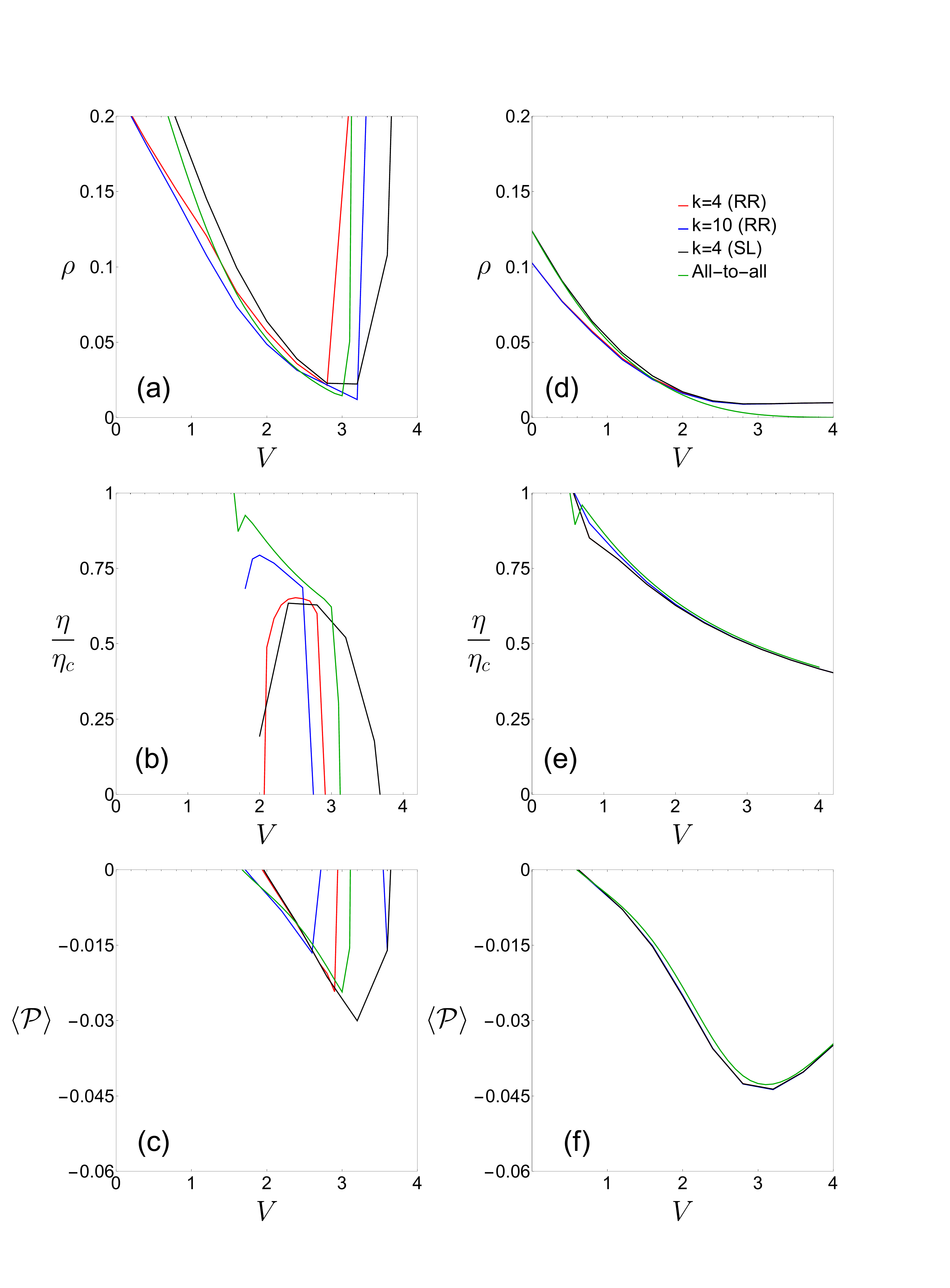} 
               \caption{Results for homogeneous
               topologies with distinct  connectivities $k$'s. Depiction   of density $\rho$ (top),  efficiency $\eta/\eta_c$ (center) and $\langle {\cal P}\rangle\equiv {\cal P}/N$ (bottom) versus coupling for $\epsilon=0.1$ (left)
               and $1$ (right).   {Symbols RR and SL denote square-lattice and random-regular topologies, respectively.} Parameters: $\beta_1=5,\beta_2=1,F=1$ and $E_a=2$.}
        \label{fig2x}
    \end{figure}

\begin{figure}
            \includegraphics[width=.56\textwidth]{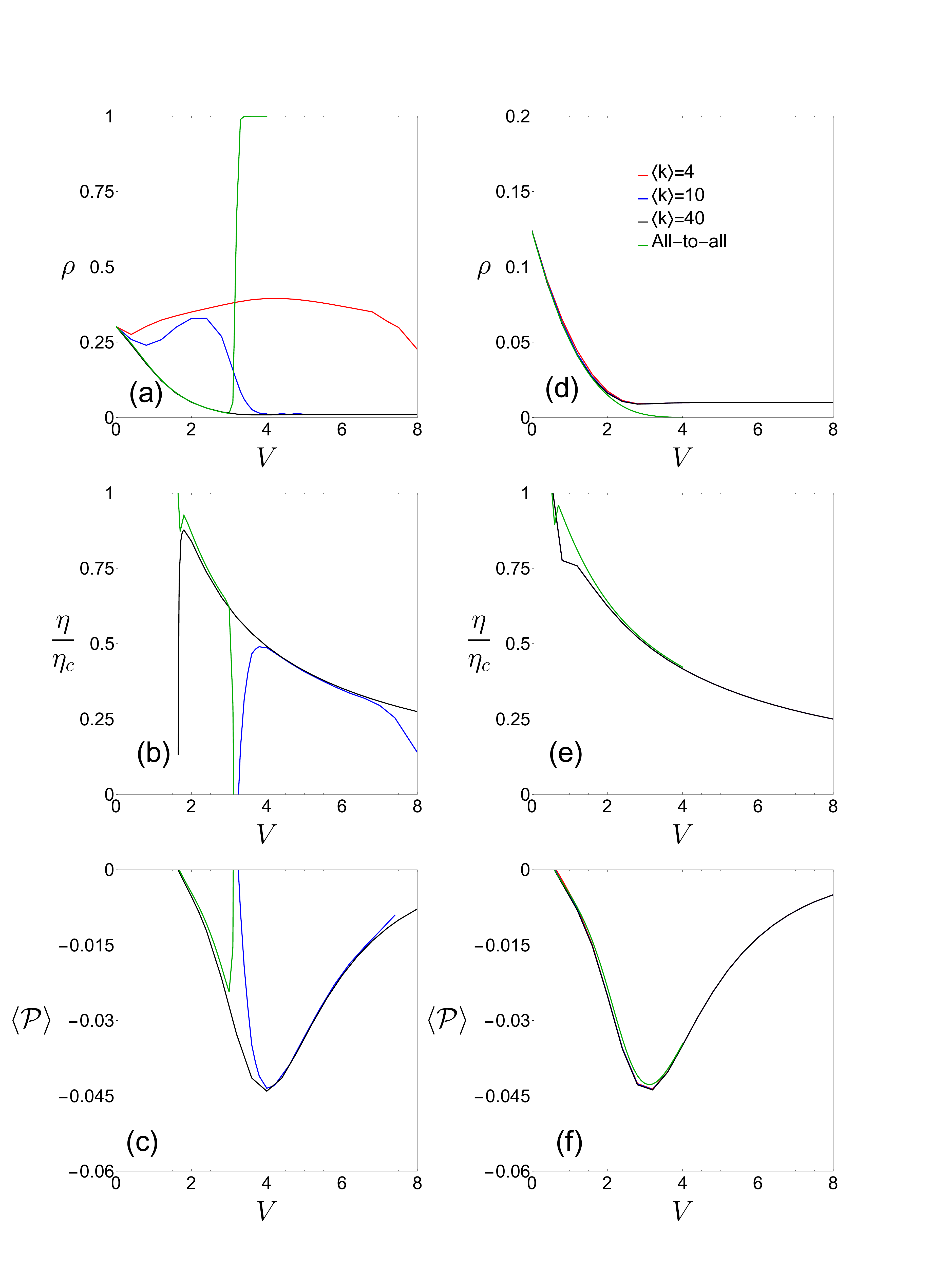} 
               \caption{Results  for heterogeneous
               topologies with distinct mean connectivities $\langle k\rangle$. Depiction   of $\rho$ (top), efficiency $\eta/\eta_c$ (center) and $\langle {\cal P}\rangle\equiv {\cal P}/N$ (bottom) versus coupling for $\epsilon=0.1$ (left)
               and $1$ (right). Parameters: $\beta_1=5,\beta_2=1,F=1$ and $E_a=2$. }
        \label{fig3x}
    \end{figure}

In this section, we will go beyond the
minimal models and look at both homogeneous and heterogeneous structures. 
Unlike minimal models, it is not possible to obtain analytical expressions and our analysis will focus on numerical simulations using the Gillespie method \cite{gillespie1977exact}. 
Due to the existence of several parameters ($\beta_1,\beta_2,F,\epsilon,V$
and $E_a$), we shall center our analysis on $F=1,\beta_1=5$ and $\beta_2=1$, in which results minimal models  predict a marked  engine regime
as $V$ is varied.
Fig. \ref{fig2x} and  \ref{fig3x} depict some results for homogeneous and heterogeneous structures, respectively.

Starting our analysis for $\epsilon=0.1$ (top panels) and homogeneous arrangements, we see [Fig.~\ref{fig2x}(a) and (c)] that the  system
performance increases by increasing the connectivity $k$ and there are small differences between regular and random-regular arrangements. Unlike the homogeneous case, differences
between $\langle k\rangle$'s are 
particularly clear for heterogeneous structures, where the heat engine is absent for $\langle k\rangle=4$. In this case, the system only operates as a pump, similarly to the stargraph, see e.g. Fig. \ref{fig32} for $N=20$. On the other hand, the heat engine is present for $\langle k\rangle=10$ and $40$. A possible explanation
is that the former and latter cases are closer to the stargraph and the all to all structures, respectively.

The results for $\epsilon=1$ (bottom panels) are remarkably different. We see that  the heat engine regime becomes much larger in terms of $V$ (with $\rho$ monotonously decreasing as $V$ goes up). Furthermore, one can see that { the influence of the lattice topology and neighborhood becomes negligible and the results become very similar to those of the all-to-all topology, revealing that
the role of topology is not so important for larger $\beta_\nu\epsilon$'s}.
%In the last analysis, we compare the performance of all studied structures %together.  For that, we shall focus our analysis on the following set of %parameters: $\beta_1=1,\beta_2=0.5,F=1,\epsilon=0.1$, whose results are %shown xxxx
%\begin{figure}
%            \includegraphics[scale=0.32]{fig7.png} 
%               \caption{For distinct lattice topologies, the comparison %among efficiency and power-outputs curves for $\epsilon=0.1$ (left) and %$\epsilon=0.5$ (right). Parameters: $\beta_1=5,\beta_2=1,F=1$ and $E_a=2$. %}
%        \label{fig7}
%    \end{figure}
%A global comparison among all structures in such case is depicted in Sec. \ref{app}.

\section{conclusions }\label{Conclusions}

%The performance of nonequilibrium engineered setups have been commonly investigated by changing the drivings, temperature of thermal baths and model parameters. 
In this paper, we studied the role of  topology of interactions on the performance of thermal engines. We investigated four distinct topologies for a simple setup composed of {interacting  unicyclic machines, each one allowed to be in two states}:  all-to-all, stargraph, homogeneous and heterogeneous structures. Different findings can be extracted from the present study. Interestingly, the  interplay among parameters
(individual $\beta_\nu\epsilon$, interaction energies $V$ and temperatures) provides two opposite scenarios, in which the role of topology is important and less important respectively, depending on whether  {$\beta_\nu\epsilon$} is small or large. The former case not only shows  a discontinuous phase transition as the interaction
is raised, but also how  the increase of neighborhood (both homogeneous and heterogeneous)
increases the efficiency but in contrast its power is inferior.  Since a majority fraction of  are empty in the latter case, the topology of interactions plays no major role.

As a final comment, we mention some ideas for future research. It might be interesting to study the the full statistics of power and efficiency in different lattice topologies, in order to tackle the  influences of fluctuations. Also, it might be interesting to compare the performance of different engine projections, such as those composed interacting units  placed in contact with only one thermal bath in {instead} of two, in order to compare the system's performances as well a the influence of lattice topology in those cases. {Finally,
it shall be interesting to investigate the inclusion of interactions between
units in the same sate (as considered in Ref.~\cite{PhysRevResearch.5.023155})
as its competition with interactions given by Eq.~(\ref{interac}).}

\section{Acknowledgments}We acknowledge the financial support from Brazilian agencies CNPq and FAPESP under grants 2021/03372-2, 2021/12551-8 and 2023/00096-0.

\bibliography{refs}

\onecolumngrid
\begin{center}
\textbf{\large Appendix}
\end{center}
\setcounter{equation}{0}
\setcounter{figure}{0}
\setcounter{table}{0}
\setcounter{page}{1}
\setcounter{section}{0}
\makeatletter
\renewcommand{\theequation}{A\arabic{equation}}
\renewcommand{\thefigure}{A\arabic{figure}}
\renewcommand{\citenumfont}[1]{A#1}

\subsection{Power-output heat maps for the minimal models}\label{app}
\begin{figure}[h]
    \centering
    \includegraphics[scale=0.4]{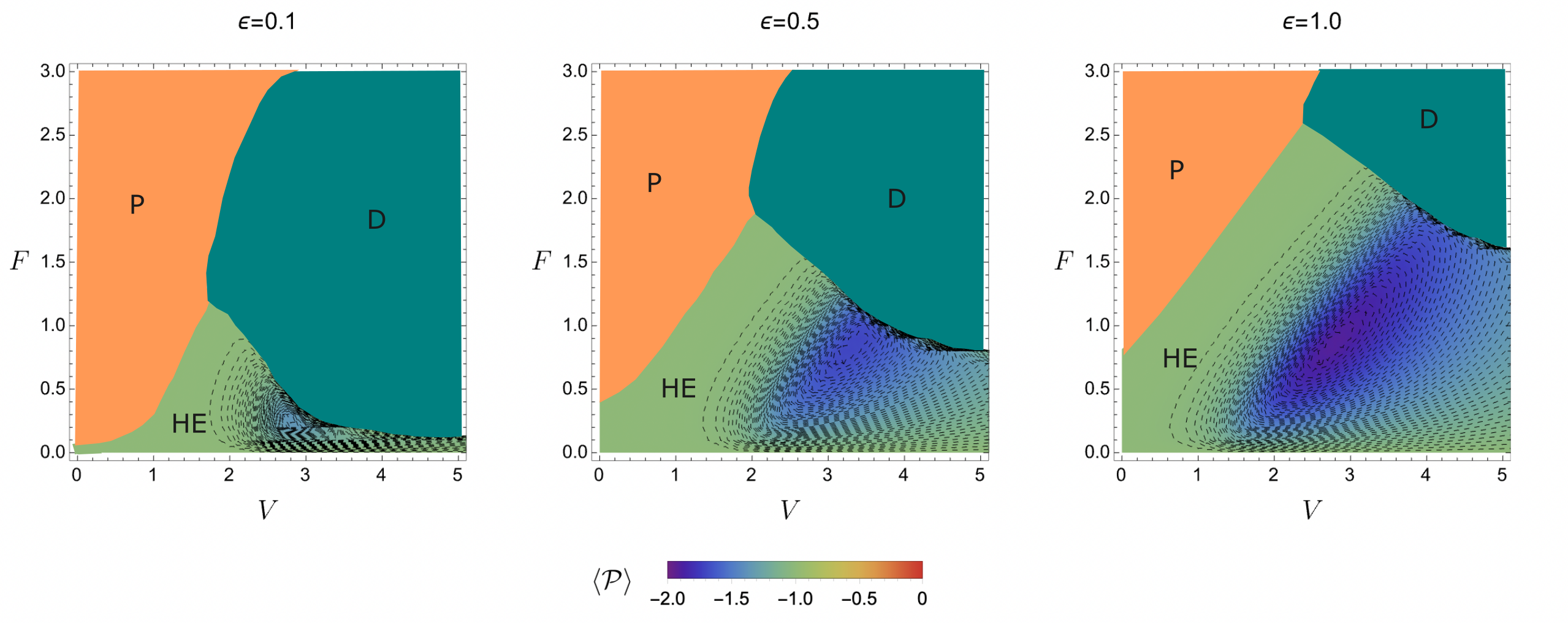}
    \caption{For the all-to-all case, $\mom{\mathcal{P}}$ heat maps for
    the same $\epsilon$ in Fig. \ref{fig32}. Parameters: $N=20$, $E_a=2$, $\beta_1=10$, $\beta_2=1$.}
    \label{figa1}
\end{figure}

\begin{figure}[h]
    \centering
    \includegraphics[scale=0.4]{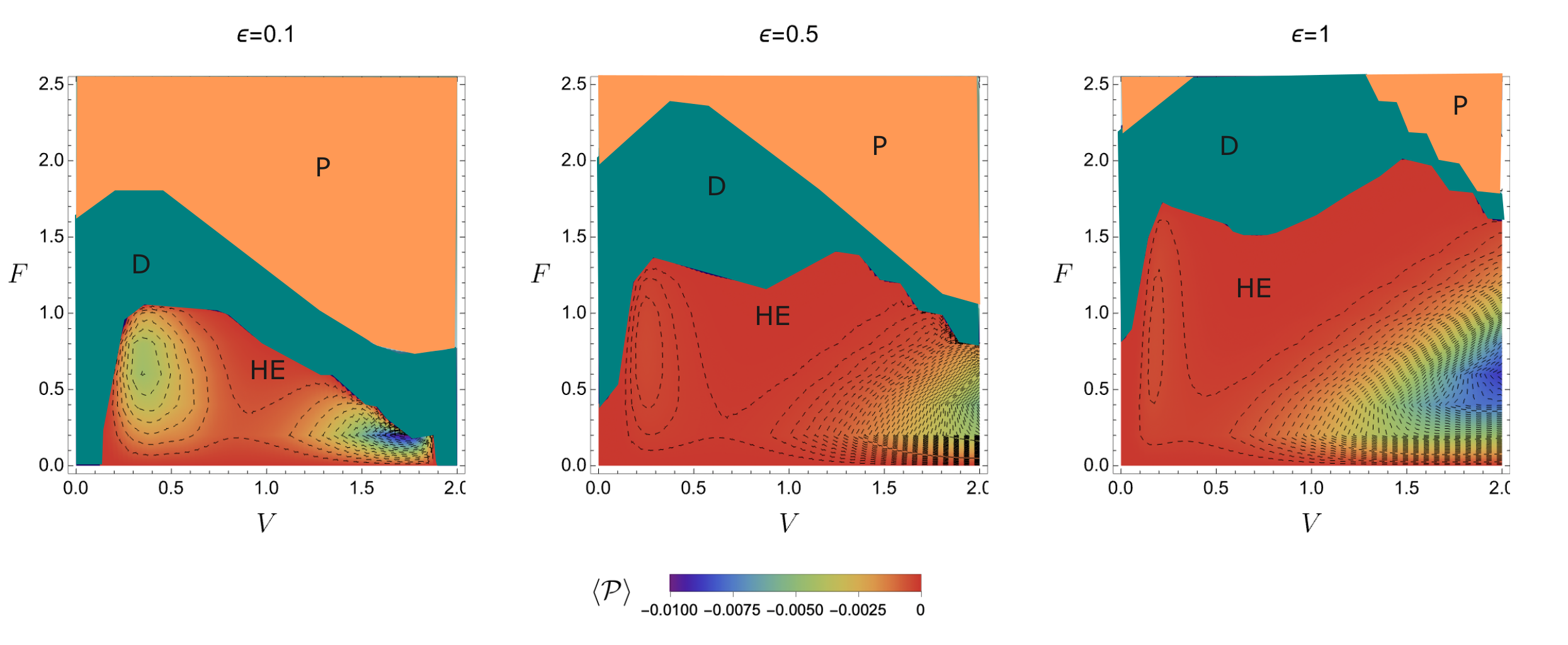}
    \caption{Heat maps for the power of the stargraph model for different values of $\epsilon$.Parameters: $N=20$, $E_a=2$, $\beta_1=10$, $\beta_2=1$}
    \label{figa2}
\end{figure}

In this appendix, we show in Figs.~\ref{figa1} and \ref{figa2} the heat maps for the power output for both all-to-all and stargraph cases for $N=20$ for the same parameters from figs. 4 and 5. 

\begin{figure}[h]
    \centering
    \includegraphics[scale=0.4]{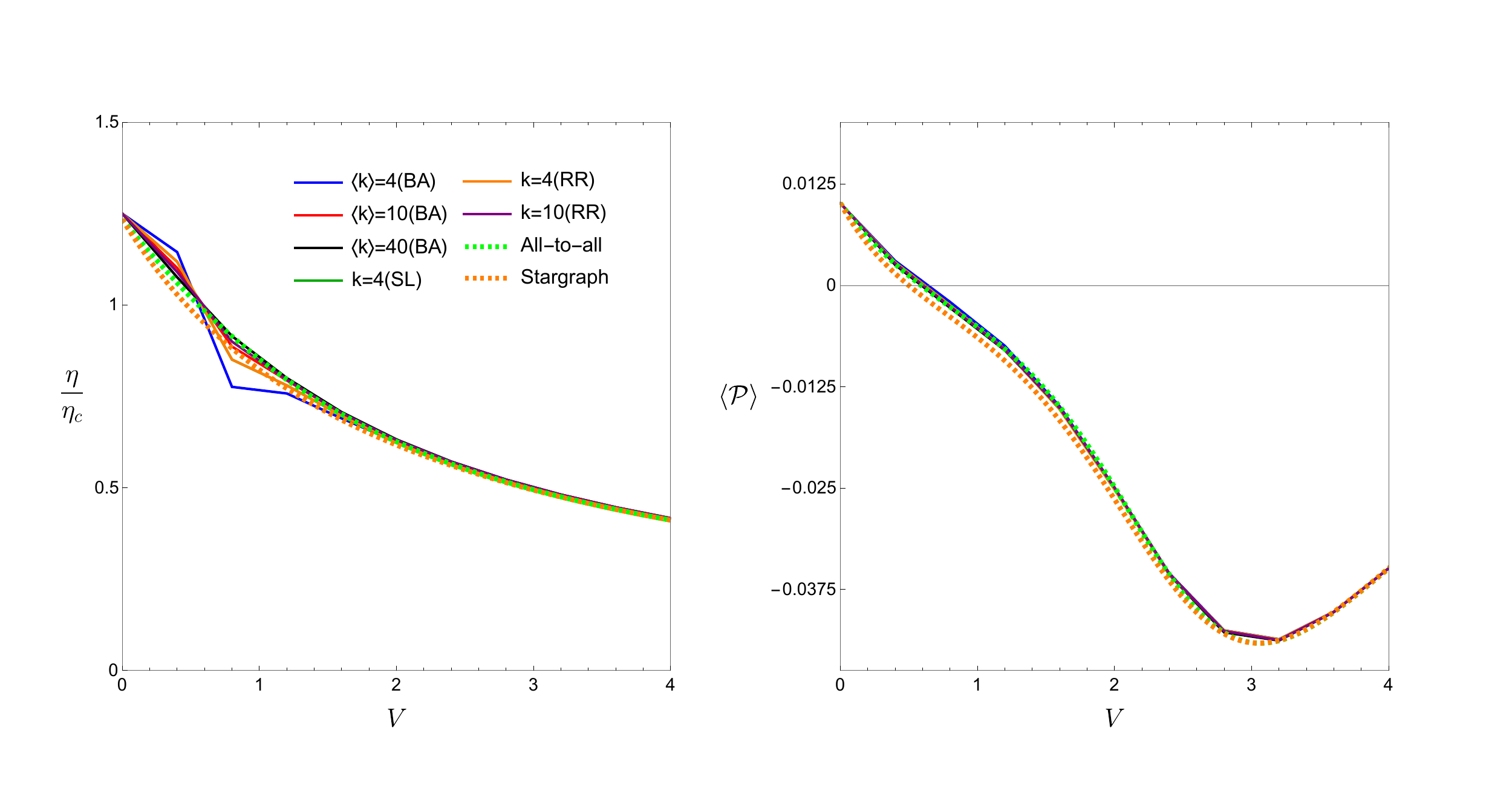}
    \caption{Depiction of $\eta/\eta_c$ and $\mom{\mathcal{P}}$ for all the topologies. {Symbols RR, SL and BA denote square-lattice, random-regular and heterogeneous (Barabasi-Albert) topologies, respectively.}Parameters:$\beta_1=5, \beta_2=1, E_a=2, \epsilon=1$ and $F=1$.}
    \label{figa3}
\end{figure}

Finally, Fig.~\ref{figa3} draws a global comparison among all structures for $\epsilon=1$. As can be seen, there is small
difference among structures, conferring some somewhat
superior efficiencies for large connectivities.

\end{document}